\documentclass[%
onecolumn,
 amsmath,amssymb,
 aps,
 prl,
]{revtex4}
\usepackage{dcolumn}
\usepackage{bm}
\usepackage{graphicx}
\usepackage{hyperref}
\usepackage{comment}
\usepackage{color}
\usepackage{soul}
\usepackage{pdfpages}

\begin{document}


\title{ 
Optimal mean first-passage time for a Brownian searcher subjected  to
resetting: experimental and theoretical results}

\author{Benjamin Besga$^1$, Alfred Bovon$^1$, Artyom Petrosyan $^1$, Satya N. Majumdar $^2$, Sergio 
	Ciliberto $^1$}\email[E-mail me at: ]{sergio.ciliberto@ens-lyon.fr}

\affiliation{$^1$ Univ Lyon, Ens de Lyon, Univ Claude Bernard, CNRS, 
Laboratoire de Physique, UMR 5672, F-69342 Lyon, France}

\affiliation{$^2$ LPTMS, CNRS, Univ. Paris-Sud, Universit\'e 
Paris-Saclay, UMR 8626, 91405 Orsay,	France}

\date{\today}

\begin{abstract}

We study experimentally and theoretically the optimal mean time needed by a free diffusing Brownian 
particle to reach a target at a distance $L$ from an initial position in the presence of resetting. 
Both the initial position and the resetting position are Gaussian distributed with width $\sigma$.  We 
derived and tested two resetting protocols, one with a periodic and one with random (Poissonian) 
resetting times. We computed and measured the full first-passage probability 
distribution that displays spectacular spikes immediately after each resetting time for close targets. 
We study the optimal mean first-passage time as a function of the resetting 
period/rate for different values of the ratio $b=L/\sigma$ and find an interesting phase transtion at a 
critical value $b=b_c$. For $b_c<b<\infty$, there is a {\it metastable} optimum time which disappears 
for $b<b_c$. The 
intrinsic diffculties in implementing these protocols in experiments are also discussed.

\end{abstract}

\maketitle


When searching for a lost object in vain for a while, intuition tells us
that maybe one should stop the current search and restart the search
process all over again. The rational behind this intuition is that
a restart may help one to explore new pathways, thus facilitating
the search process. This intuition has been used empirically before
in various stochastic search algorithms (such as simulated annealing)
to speed up the search process~\cite{VAVA91,LSZ93,TFP08,Lorenz18}.
More recently, in the physics literature, this fact was demonstrated 
explicitly by studying the mean first-passage
time (MFPT) for a single particle (searcher) to a fixed target in various models, in
the presence of resetting (for a recent review see ~\cite{reset_review}).
 
Searching a target via resetting is an example of the so called
intermittent search strategy~\cite{BLMV11} that consists of a mixture of
short-range moves (where the actual search takes place) with
intermittent long-range moves where the searcher relocates to
a new place and starts a local search in the new place. 
More precisely, let there be a fixed target at some point in space
and a particle (searcher) starts its dynamics from a fixed initial position
in space. The dynamics of the particle may be arbitrary, e.g., it
may be simply diffusive~\cite{EM2011_1,EM2011_2}.
Resetting interrupts
the natural dynamics of the particle either randomly at a constant 
rate $r$~\cite{EM2011_1,EM2011_2}
or periodically with a period $T$~\cite{PKE16,BBR16}, and sends 
the particle to its initial
position. The dynamics starts afresh from the initial position
after each resetting event. The MFPT to find
a target, located at a fixed distance away from the initial position,
typically shows a unique minimum as a function of $r$ (or $T$ for periodical
resetting).
These studies have led to the paradigm that resetting typically makes
the search process more efficient and moreover,
there often exists an optimal resetting rate $r^*$
(or period $T^*$) that makes the search time minimal~\cite{EM2011_1,EM2011_2}.

While this `optimal resetting' paradigm has been tested and verified in
a large number of recent theoretical and numerical 
studies~\cite{EMM13,MV13,EM2014,KMSS14,RUK14,RRU15,CS15,KGN15,PKE16,NG16,Reuveni16,PR17,CS18,Belan18,EM18,MPCM19a,MM19,DLLJ19,PP19} 
(see also the review \cite{reset_review}), 
to the best of our knowledge 
it has not been verified experimentally (see however the very recent preprint~\cite{Tal2020}, 
which appeared in arXiv during the writing of this article, where the authors used
a holographic optical twizzer set-up. Their resetting protocols though are quite different
from ours).
The purpose of this Letter is to report
an experimental realization using optical laser traps and 
subsequent test/verification of this 
`optimal resetting' paradigm.
The goal of the experiment is not just to mimic the theoretical models,
but we will see that designing an experiment with a realistic resetting
protocol is challenging. Moreover, our experimental protocol led us, in turn,
to study new models that exhibit interesting, rich and novel 
phenomena associated to resetting, namely metastability and phase transition
in the MFPT as a function of resetting rate/period.

\begin{figure}[htbp]
\includegraphics[width=0.45\textwidth]{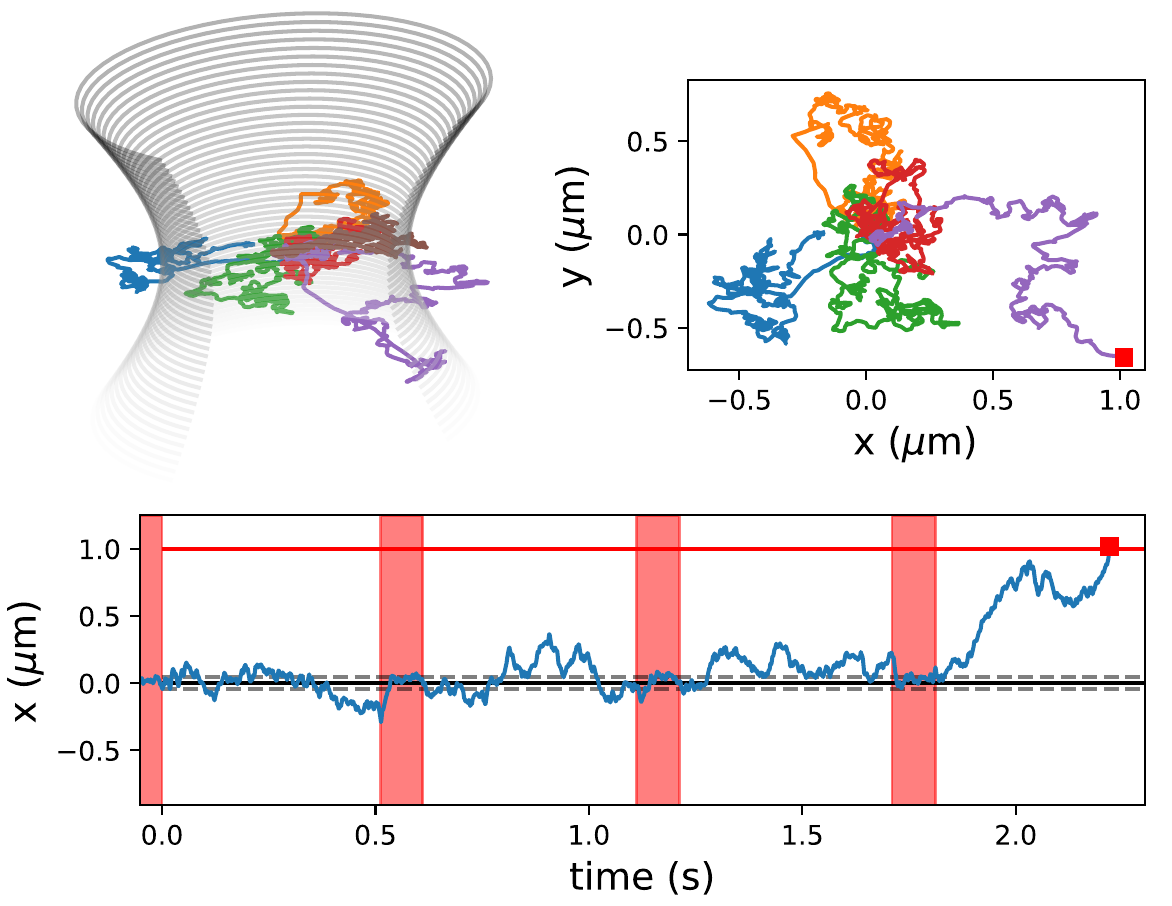} 
\caption{\textit{Top panel :} Sketch of the optical trap and a Brownian trajectory in the 2D plane (x,y) with resetting.
\textit{Bottom panel :} typical 1D Brownian trajectory with periodic resetting (red colored area) 
at time $T = 0.5$ s. The equilibrium standard deviation of the trapped particle $\sigma = 43 $ nm is shown 
(dotted lines). The target (red square 1 $\mu$m away from the center of the trap) is reached 
when $x$ crooses the red line for the first time.}
\label{fig_setup} 
\end{figure}

{\it Experimental set-up.} We have implemented an optimal search protocol with a Brownian particle which is 
periodically or randomly reset using an optical tweezers \cite{Berut2016}. We use silica micro-sphere of radius 
$R=1$ $\mu$m ($\pm 5 \%$) immersed in pure water. The fluid chamber is designed to have very few particles in the 
measuring volume which allows us to take long measurements. A near infrared laser 
with wavelength $\lambda = 1064$ nm is 
focused into the chamber through an oil immersion objective (Leica $63\times$ and $1.4$ NA) to create an optical 
trap. The particle is trapped in two dimensions in the (x,y) plane by a harmonic potential with stiffness 
$\kappa$. The stiffness is controlled by changing the optical power in the chamber directly by laser current 
modulation (up to 10 kHz) or by means of an electro-optical modulator (EOM).

The position of the particle in the (x,y) plane can be tracked (see Fig. \ref{fig_setup}) either by a 
white light imaging on a camera (maximum speed around 1000 frames per second) or from the deviation of a red 
laser on a quadrant photodiode (QPD) with a band pass around 1 MHz. 
The experiments were performed close to the bottom surface of the cell to avoid errors in position 
reading due to sedimentation. Henceforth, for simplicity, we
will focus only on the $x$-component of the particle's position, i.e., the one dimensional
trajectory. We can thus follow the particle position during its free diffusion and after a time 
$T$ reset its position by 
turning on the optical trap which allows us to realize different protocols.

{\it Periodic resetting protocol.} The experimental protocol leads us to study
a model of diffusion in (effectively) one dimension subjected to periodic resetting.
We consider  an overdamped particle in thermal equilibrium
inside a harmonic trap with potential $U(x)= \kappa\, x^2/2$, 
where the stiffness $\kappa$ is proportional to the trapping laser power. 
This means that the initial position
$x_0$ is distributed via the Gibbs-Bolzmann distribution which is
simply Gaussian: ${\cal P}(x_0)= e^{-x_0^2/2\sigma^2}/\sqrt{2\pi \sigma^2}$
where the width $\sigma= \sqrt{k_B T/\kappa}$. We also consider a fixed target
at location $L$. 
At time $t=0$, the trap
is switched off for an interval $T$ and the particle undergoes free diffusion
(overdamped) with diffusion constant $D=k_B T/\Gamma$ (here $\Gamma$
denotes the damping constant). At the end of the period, the particle's 
position is reset (see Fig. \ref{fig_setup}). Performing the resetting of the
position poses the real experimental challenge. 

In standard models of resetting one usually assumes 
instantaneous resetting~\cite{reset_review}, which is however
impossible to achieve experimentally.
There have been recent theoretical studies that incorporates a `refraction'
period before the particle's position is reset~\cite{Reuveni16,EM2019,PKR2019,GPP2019}. 
In our experiment, we have an analogue of this refraction period: after 
the period $T$, we switch on the optimal harmonic trap
and we let the particle relax back to its equilibrium thermal distribution.
This relaxation can, in principle, be made
arbitrarily fast using, e.g., the recently developed
`Engineered Swift Equilibration' (ESE) technique~\cite{ESE1,ESE2,RMP19,PGTP19,GPP19}.
In our experiment we determine the characteristic relaxation time $\tau_c$ and during the period
$[T, T+\tau_{\rm eq}]$ with $\tau_{\rm eq} \simeq 3 \tau_c$, we do not make
any measurement. In other words, 
even if the particle encounters the target during the relaxation period, 
we do not count that event as a first-passage event. Thus in this  
set-up, the thermal relaxation mimics the instantaneous resetting, with
one major difference however. We do not reset it to exactly 
the same initial position, rather the new `initial' position at 
the end of the time epoch $T+\tau_{\rm eq}$ is drawn from the Gibbs-Boltzmann
distribution ${\cal P}(x_0)$. At time $T+\tau_{\rm eq}$, we again switch off the trap
and we let the particle diffuse freely for another period $T$, followed by 
the thermal relaxation over period $\tau_{\rm eq}$. The process repeats periodically.
During the free diffusion, if the particle finds the target at $L$, we
 measure the first-passage time $t_f$. Note that the 
first-passage
time $t_f$ is the net
`diffusion' time spent by the particle before reaching the target (not counting the intermediate relaxation periods $\tau_{\rm eq}$).
Averaging over many realizations,
we then compute the MFPT $\langle t_f\rangle$, for fixed
target location $L$ and fixed resetting period $T$.

\begin{figure}[t]
\includegraphics[width=0.5\textwidth]{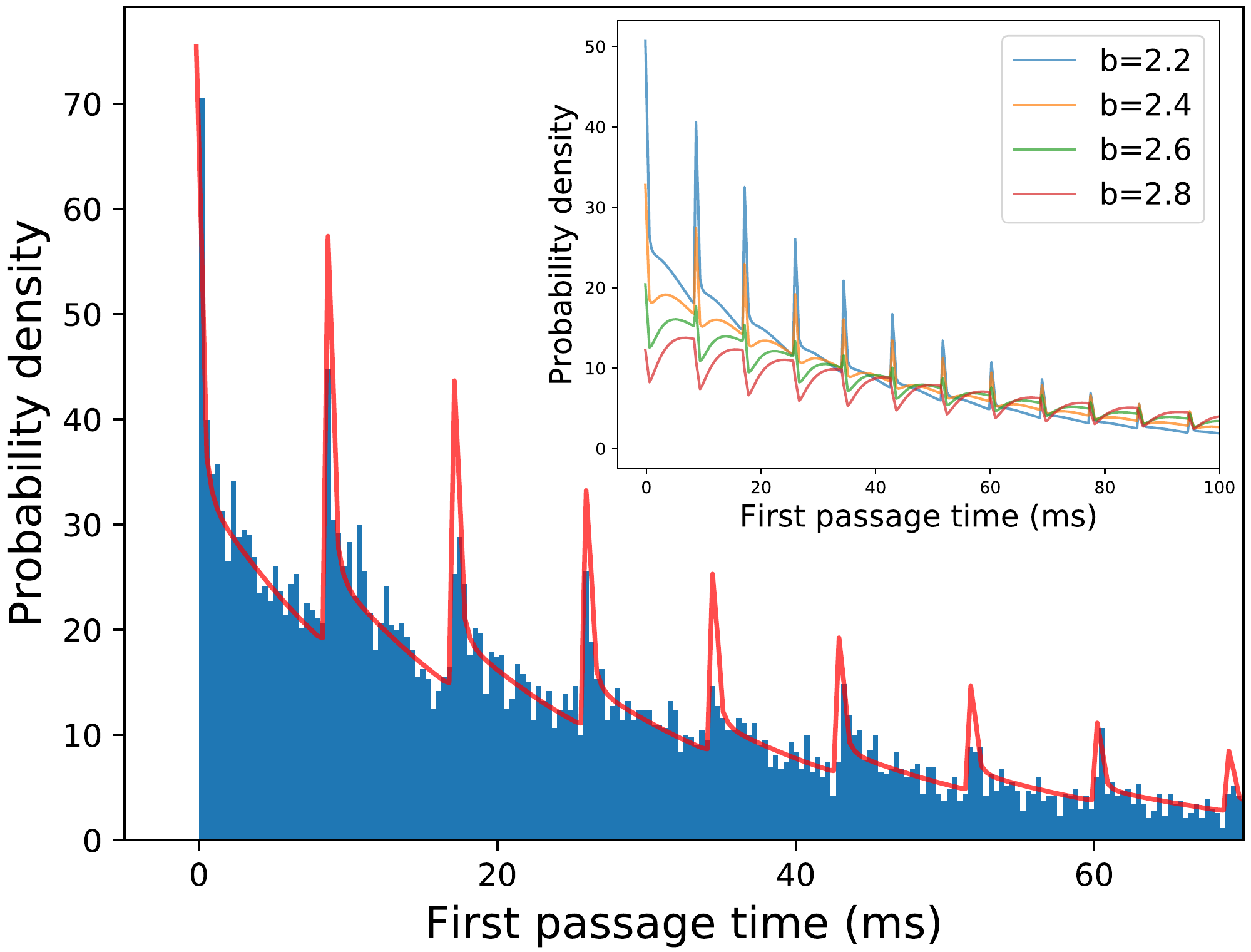} 
\caption{Experimental (blue histogram) and theoretical (red line) PDF $F(t)$ of the 
first-passage time. We measured $1.2 \times 10^4$ first-passage times for $b=2$ and $c=1$ ($T=8.6$ ms).
The spikes occur just immediately after $t=n\, T$ where $n=0, \, 1,\,  2, \ldots$. 
\textit{Inset:} The PDF $F(t)$ vs. $t$ (theoretical) for $b = 2.2$ to $2.8$ that demonstrates
that the spikes disappear rapidly with increasing $b$. 
The PDF $F(t)$ is normalized such that $\int_0 ^\infty F(t) dt = 1$ with  
$dt = 3.5\times 10^{-4}$ s.}
\label{fig_stat} 
\end{figure}

The MFPT for this protocol can be computed exactly, as detailed in the Supp. Mat.~\cite{supp_mat}. Our 
main result can be summarized in terms of two dimensionless quantities \begin{equation} b= 
\frac{L}{\sigma}\, ; \quad {\rm and} \quad c= \frac{L}{\sqrt{4\,D\,T}}\, . \label{bc_def} \end{equation} 
The parameter $b$ quantifies how far the target is from the center of the trap in units of the trapped 
equilibrium standard deviation. The second parameter $c$ tells us how fast we reset the particle 
compared to its free diffusion time. We show that the dimensionless MFPT $\tau$ can be expressed as a function 
of these two parameters $b$ and $c$ 
\begin{equation} 
\tau=\frac{4D\langle t_f\rangle}{L^2}= w(b,c)\, , 
\label{tau_bc} 
\end{equation} 
where 
\begin{equation} 
w(b,c)=\frac{\int_0^1 dv \int_{-\infty}^{\infty} 
du\, e^{-u^2/2}\, {\rm erf}\left( \frac{c}{\sqrt{v}}\,|1-u/b| \right)}{c^2\, \int_{-\infty}^{\infty} 
du\, e^{-u^2/2}\, {\rm erfc}\left(c\,|1-u/b|\right)}\, . 
\label{wbc_def} 
\end{equation} 
While it is hard to evaluate the integrals explicitly, $w(b,c)$ can be easily plotted numerically 
to study its dependence on $b$ and $c$, as discussed later. Going beyond the first moment
$\langle t_f\rangle$ and computing the full probability density function (PDF) 
of the first-passage time $t_f$
is also of great interest~\cite{pers_current,Redner,Louven_review,pers_review}.
Indeed, we computed the PDF $F(t)$ of $t_f$
(see ~\cite{supp_mat}) and the result is plotted in Fig. (\ref{fig_stat}). For small $b$, we found
striking spikes in $F(t)$ just after each resetting event. We show in ~\cite{supp_mat} that
setting $t=n\, T+ \Delta$ with $n=0,\,1,\,2\ldots$ and $\Delta\to 0^+$, 
the first-passage density $F(t=n\, T+\Delta)$ displays
a power law divergence (the spikes) as $\Delta\to 0^+$
\begin{equation}
F(t=n\, T+ \tau)\simeq A_n(b)\, \Delta^{-1/2}
\label{spike.1}
\end{equation}
with an amplitude $A_n(b)$ that can be computed explicitly~\cite{supp_mat}. 
We find that as $b$ increases, $A_n(b)$
decays rapidly and the spikes disappear for large $b=L/\sigma$ (see the inset of Fig.  (\ref{fig_stat})).
Instead for large $b$, $F(t)$ drops by a finite amount after each period (as seen in the inset).  
In theoretical models of resetting to a fixed initial position
($\sigma=0$ or $b\to \infty$), these spikes are completely absent and hence they constitute clear 
signatures of the finiteness of $\sigma$. Physically, these spikes occur for finite $\sigma$ because
each resetting event enhances the probability to find the target without the need of diffusion.

In order to test experimentally these results we realized a periodic resetting protocol and measure the 
statistics of first-passage times. The diffusion coefficient (typically $D \simeq 2.10^{-13} \,{\rm m}^2/{\rm s} $) is measured during the free diffusion 
part and the width of the Gaussian (typically $\sigma \simeq 40$ nm) when the particle is back to equilibrium. 
These independent and simultaneous measures allow us to overcome experimental drifts. 
In figure \ref{fig_stat} we show the experimentally obtained PDF of $10^4$ measured 
first-passage times for $b=2$ and $c=1$. We also compared with
our theoretical prediction~\cite{supp_mat}. 
We observe a very good agreement with no free parameter. 

\begin{figure}[htb]
\includegraphics[width=0.45\textwidth]{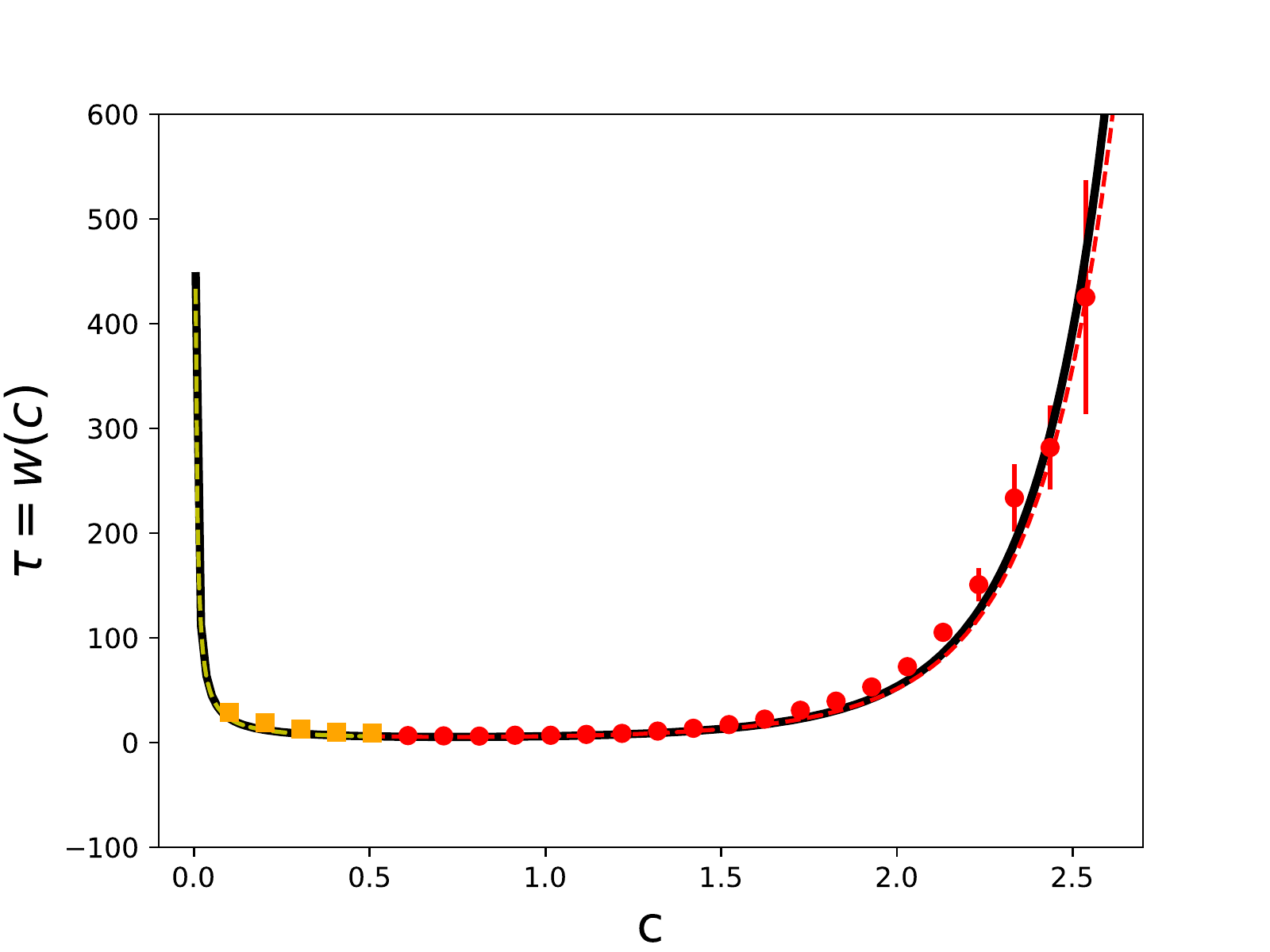} 
\caption{$w(c)\equiv w(b\to \infty,c)$ vs. $c$ curve. The solid black line represents the
theoretical formula for $w(c)$ in Eq. (\ref{delta_reset.1}), while the dots denote
the experimental data with $b = 30$ (red dots) and $b=5$ (yellow square). The dotted lines shows 
$w(b,c)$ with $b=5$ for small $c$ in yellow, and with $b=30$ for higher $c$ in red. 
The error bars are given by the standard deviation of the MFPT distribution 
divided by the square root of the number of events.}
\label{fig_wc.1} 
\end{figure}

To analyse the MFPT, we start with the limit $b\gg 1$ of Eq. (\ref{wbc_def}), i.e., $L\gg \sigma$. This 
limit corresponds to the case when the target is much farther away 
compared to the
typical fluctuation of the initial position. In this case, taking
$b\to \infty$ limit in Eq. (\ref{wbc_def}) we get
\begin{equation}
w(c)=w(b\to \infty,c)= \frac{{\rm
erf}(c)+2c\left(e^{-c^2}/\sqrt{\pi}-c\, {\rm erfc}(c)\right)}{c^2\,
{\rm erfc}(c)}\, ,
\label{delta_reset.1}
\end{equation}
where ${\rm erf}(c)= (2/\sqrt{\pi})\, \int_0^c e^{-u^2}\, du$ and
${\rm erfc}(c)=1- {\rm erf}(c)$. In Fig.~(\ref{fig_wc.1}), we plot
$w(c)$ vs $c$ and compare with our experimental data and find good 
agreement with no adjustable parameter.
Typically a few thousands of first-passage 
events (several hours of measurements) are needed to have a good estimate of the MFPT 
given the slowly decaying distribution of the 
first-passage times. The standard deviation of first-passage times is indeed of the same order as the MFPT. 
We see a distinct optimal value around $c^*=0.74$,
at which $w(c^*)=5.3$. Our results thus provide a clear 
experimental verification of this optimal resetting paradigm.
Let us remark that the authors
in Ref.~\cite{PKE16} studied
periodic resetting to the fixed initial position $x_0=0$
and obtained the MFPT by a different method than ours.
Our $b\to \infty$ limit result in Eq. (\ref{delta_reset.1}) indeed
coincides with that of ~\cite{PKE16}, since when $\sigma\ll L$, our protocol
mimics approximately a resetting to the origin.

\begin{figure}[h]
\includegraphics[width=0.45\textwidth]{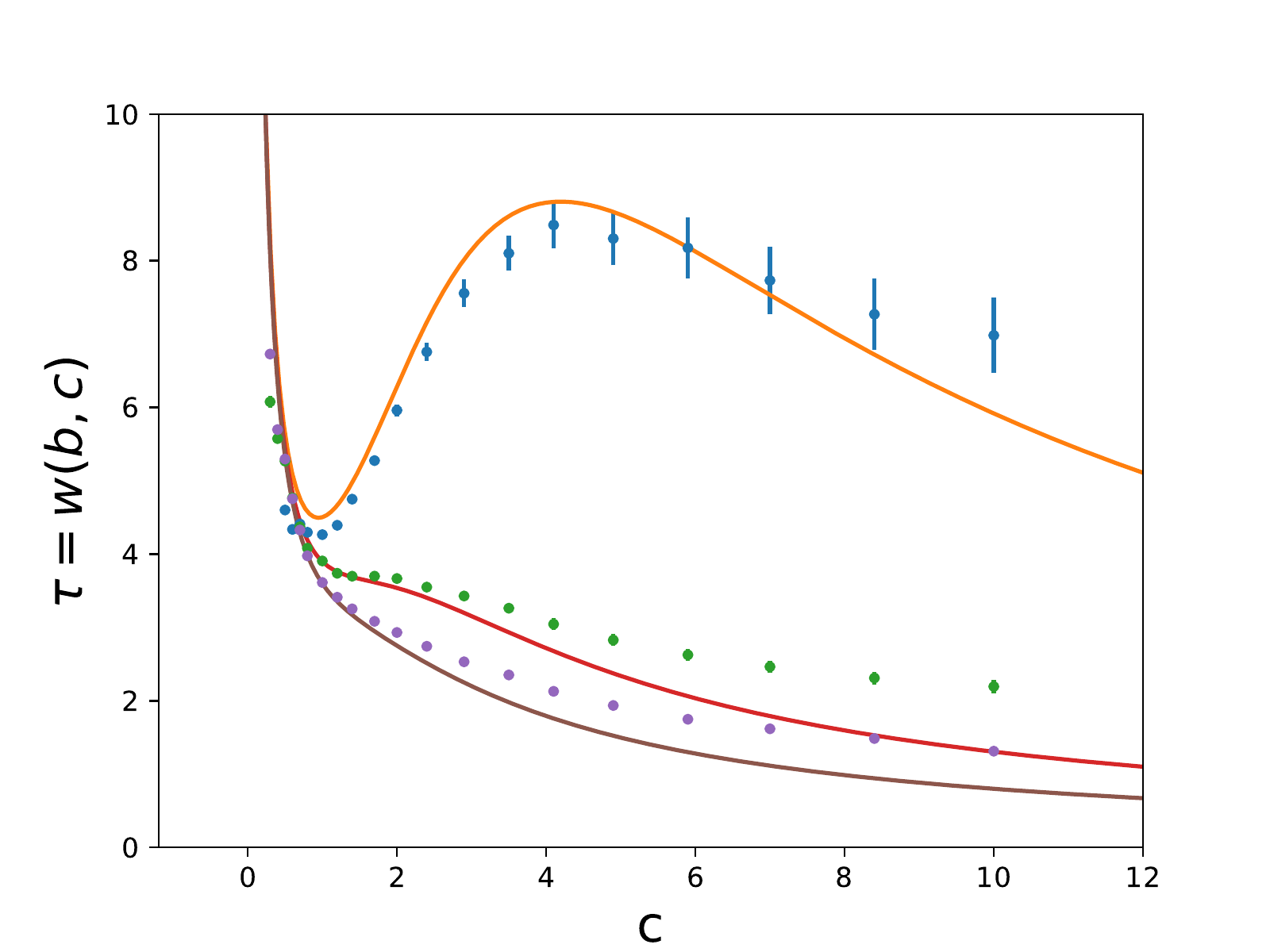}
\caption{The scaled MFPT $\tau = w(b, c)$ vs. $c$ curves and experimental 
data for $b=3$ (orange curve and blue dots), $b=2.3$ (red curve and green dots) 
and $b=2$ (brown curve and purple dots), for the periodic resetting protocol. 
The theoretical curves are obtained from Eq. (\ref{wbc_def}).
The scaled MFPT has two different behavior depending on how far is the target. If $b > b_c\approx 
2.3$ the MFPT $\tau$ exhibits a local minimum, whereas it decreases monotonically for $b$ lower than the 
critical value $b_c$. }

\label{fig_wbc}
\end{figure}

What happens when $b=L/\sigma$ is finite? Remarkably, when we plot $w(b,c)$
in Eq. (\ref{wbc_def}) as a function of $c$ for different fixed values of $b$,
we see that $w(b,c)$ decreases as $c$ increases, indeed achieves a minimum 
at $c_1(b)$, then increases, achieves a maximum at $c_2(b)$ and then decreases
monotonically as $c$ increases beyond $c_2(b)$ (see Fig. (\ref{fig_wbc})).
When $b\to \infty$, the maximum at $c_2(b)\to \infty$ and one has a
true minimum.
However, for finite $b$, the ``optimal'' (minimal) MFPT at $c=c_1$ 
is thus actually a metastable
minimum and the true minimum occurs at $c\to \infty$, i.e., when the
resetting period $T\to 0$. Physically, the limit $T\to 0$ corresponds to repeated (almost
continuously) resetting and since the target position and the initial location
are of the same order, the particle may find the target simply by resetting,
without the need to diffuse. Interestingly, this metastable minimum
exists only for $b>b_c \approx 2.3$. When $b<b_c$, the curve $w(b,c)$
decreases monotonically with $c$ and there is only a single
minimum at
$c\to \infty$, or equivalently for $T\to 0$. Thus the system
undergoes a ``first-order'' phase transition as $b=L/\sigma$ is tuned across
a critical value $b_c\approx 2.3$, from a
phase with a metastable optimum at a finite $c=c_1(b)$ to one where
the only minimum occurs at $c\to \infty$. 
This phase transition is well reproduced by the experimental data points. 
The deviation from theoretical prediction for high values of c is due to 
the limited experimental acquisition rate (here $50000$ Hz) that prevents us from 
detecting very fast events and thus leads to an 
overestimation of the MFPT as confirmed by our numerical simulations.
This phase transition
was rather unexpected and came out as a surprise.

\begin{figure}[h]
\includegraphics[width=0.45\textwidth]{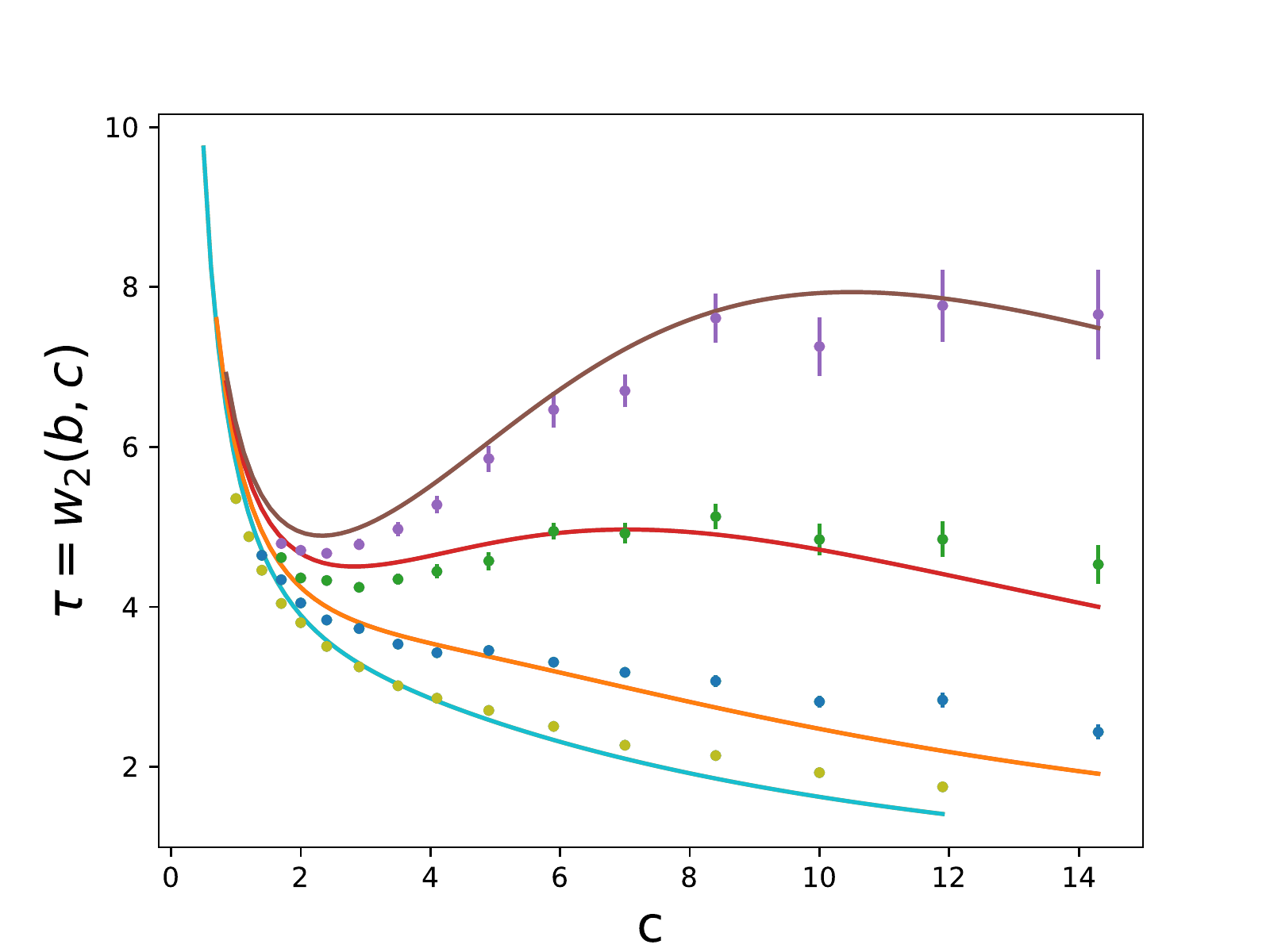}
\caption{ Scaled MFPT $\tau = w_2(b,c)$ vs $c$ curves and experimental 
data points for $b = 2, 2.3, 2.7$ and $3$ (from bottom to top)  in the case
of Poissonian resetting.
The metastable minimum disappears for $b<b_c\approx 2.53$.}
\label{fig_wbc_Poisson}
\end{figure}

{\it Random resetting protocol.}
It turns out that this metastability and the phase transition is rather
robust and exists for other protocols, such as Poissonian resetting
where the resetting occurs at a constant rate $r$. Here, the dimensionless
variables are $b=L/\sigma$ and $c=\sqrt{r/D}\, L$ 
and the scaled MFPT $\tau= 4 D\langle t_f\rangle/L^2$ again becomes 
a function $w_2(b,c)$ of $b$ and $c$  
(analogue of Eq. (\ref{wbc_def})). In this case, we get
a long but explicit $w_2(b,c)$ (see ~\cite{supp_mat} for details).
In Fig. (\ref{fig_wbc_Poisson}) we plot $w_2(b,c)$ vs. $c$
for different values of $b$ 
together with experimental data. We have a good agreement between theory and experiment and here again the 
deviation at high c comes from limited experimental acquisition rate. Once again we see that there is a 
metastable minimum that disappears when $b$ decreases below a critical value $b_c\approx 2.53$. When $b\to 
\infty$, there is only a single minimum at $c^*=1.59362$ where $w_2(\infty, c^*)= 6.17655$. Thus this 
phenomenon of metastability and phase transition in MFPT seems to be robust.

\vskip 0.3cm

{\it Conclusions :} We have studied both theoretically and experimentally the role of the variance of the 
resetting position on the optimal time needed for a Brownian bead to reach a specific immobile target. We have 
applied two different protocols, one with a periodic and another with a random resetting time. We found that 
both present a metastable minimum for $b>b_c$ (where $b_c$ depends on the protocol),
showing that this transition could be a universal feature of 
MFPT protocols for resetting. For the periodic protocol, we also computed and measured the 
full PDF of the first-passage time and found that it displays striking spikes after each resetting event,
a clear effect of the finiteness of the variance of the initial position. 
The experimental data agree well with theoretical predictions, but the experiment is 
not a mere test of the theory. Indeed we point out a series of experimental difficulties that one has to 
consider when one applies such theoretical predictions in real systems. On one side when the particle is 
free for a very long time (small $c$) there are problems of sedimentation that must be taken into account 
because they may influence a lot the final result. On the very short times (large $c$) we have shown that 
finite sampling time affects the results, because if the first-passage is not detected, it leads 
to an overestimation of the first-passage time. It is clear that at very fast sampling rate 
the model will fail and probably 
the role of inertia, which has been always neglected, should be taken into account in future theoretical 
developments. 

We thank E. Trizac for useful discussions.SNM wants to thank the warm hospitality of ICTS (Bangalore)  where this work was completed.

\newpage
\end{document}


\title{Supplementary Material of the Letter: \\ 
Optimal mean first-passage time for a Brownian searcher subjected  to
resetting: experimental and theoretical results}

\author{Benjamin Besga$^1$, Alfred Bovon$^1$, Artyom Petrosyan $^1$, Satya N. Majumdar $^2$, Sergio 
	Ciliberto $^1$}\email[E-mail me at: ]{sergio.ciliberto@ens-lyon.fr}

\affiliation{$^1$ Univ Lyon, Ens de Lyon, Univ Claude Bernard, CNRS,
Laboratoire de Physique, UMR 5672, F-69342 Lyon, France}

\affiliation{$^2$ LPTMS, CNRS, Univ. Paris-Sud, Universit\'e 
Paris-Saclay, UMR 8626, 91405 Orsay, France}

\begin{abstract}

We give the principal details of the theoretical calculations described in the main 
text of the Letter.
	
\end{abstract}

\date{\today}

\maketitle

\section{Introduction}

In this supplementary material of the Letter, we give the details on how the theoretical predictions 
plotted in Figs. 2), 3) and 4) of the main text have been obtained, both for periodic and random 
resetting protocols. As already pointed out in the main text we consider here a realistic situation 
in which the initial position at the begining of each free diffusion period (between successive 
resetting events)
has a Gaussian distribution of width $\sigma$.  The mean first-passage time (MFPT),
in the presence of resetting for either of the two protocols, was previously computed
only for the restting to a fixed initial position $x_0$~\cite{EM2011,EM2011_2,PKE16,BBR16,reset_review}. 
In practice this is not possible to realize experimentally 
because of the finite stiffness $\kappa$ of the trap which resets the particle to the initial position. 
In an optical trap at temperature $T$, the resetting position is always Gaussian distributed with a finite 
variance $\sigma^2=k_BT/\kappa$ given by equipartition. As $\kappa$ is 
proportional to the laser intensity starting with $\sigma=0$ would imply to trap with an infinite power 
which is of course not possible. Thus we study here the influence of a finite nonzero $\sigma$ on MFPT.

Consider a searcher undergoing a generic stochastic dynamics starting, say, at the initial position $x_0$.
The immobile target is located at $x=L$.
To compute the MFPT of a generic stochastic process, it is most convenient to first compute the survival probability or 
persistence~\cite{pers_current,Redner,Louven_review,pers_review} $S(t|x_0)$, starting from $x_0$.
This is simply the probability
that the searcher, starting its dynamics at $x_0$ at $t=0$, does not find the target up to time $t$. 
The first-passage probability density
$F(t|x_0)$ denotes the probability density to find the target for the first time at $t$.
The two quantities $F(t|x_0)$ and $S(t|x_0)$ are simply related to each other via $S(t|x_0)=\int_t^{\infty} F(t'|x_0)\, dt'$, 
because if the target is not found up to $t$, the first-passage time must occur after $t$. 
Taking a derivative with respect to $t$
gives
{
\begin{equation}
F(t|x_0)= - \frac{dS(t|x_0)}{dt} \, .
\label{FQ.1}
\end{equation}
Hence, if we can compute $S(t|x_0)$ (which is often easier to compute), we obtain 
$F(t|x_0)$ simply from Eq. (\ref{FQ.1}).}
Once we know $F(t|x_0)$, the MFPT is just its first moment
\begin{equation}
\langle t_f\rangle (x_0)= \int_0^{\infty} t\, F(t|x_0)\, dt= \int_0^{\infty} S(t|x_0)\, dt\, ,
\label{MFPT.0}
\end{equation}
where in arriving at the last equality, we substituted Eq. (\ref{FQ.1}) and did integration by parts. 
Below, we will first compute the survival probability $S(t|x_0)$ for fixed $x_0$
and then average over the distribution of $x_0$. We will consider the two protocols separately.

\section{Protocol-1: Periodic Resetting to a random initial position }

We consider an overdamped diffusing particle that starts at an initial
position $x_0$, which is drawn from a distribution ${\cal P}(x_0)$.
The particle diffuses for a fixed period $T$ and then its position is
instantaneously reset to a new position $z$, also drawn from the same
distribution ${\cal P}(z)$. Then the particle diffuses again for a period 
$T$, followed by a reset to a new position $z'$ drawn from ${\cal P}(z)$
and the process continues. We assume that after each resetting, the reset 
position $z$ is drawn {\rm independently} from cycle to cyle from the
same distribution ${\cal P}(z)$. We also have a fixed target at a
location $L$. For fixed $L$, $T$ and ${\cal P}(z)$, We want to 
first compute the mean first-passage time
$\langle t_f\rangle$ to find the target and then optimize (minimize) this
quantity with respect to $T$ (for fixed $L$ and ${\cal P}(z)$).
We first compute the MFPT for arbitrary ${\cal P}(z)$ and then focus on the
experimentally relevant Gaussian case.

We first recall that for a free diffusing particle, starting at an initial 
position $x_0$, the survival probability that it stays below the level $L$ up to 
time $t$ is given by~\cite{pers_current,Redner,Louven_review,pers_review}
\begin{equation}
S(t|x_0)= {\rm erf}\left(\frac{|L-x_0|}{\sqrt{4Dt}}\right)\, ,
\label{surv_diff.1}
\end{equation}
where ${\rm erf}(z)= (2/\sqrt{\pi})\, \int_0^{z} e^{-u^2}\, du$ is the
error function and $D$ is the diffusion constant. If the initial
position is chosen from a distribution ${\cal P}(x_0)$, then
the survival probability, averaged over the initial position, is given by
\begin{equation}
Q_1(t)= \int_{-\infty}^{\infty} dx_0\, {\cal P}(x_0)\, S(t|x_0)=
\int_{-\infty}^{\infty} dx_0\, {\cal P}(x_0)\, {\rm 
erf}\left(\frac{|L-x_0|}{\sqrt{4Dt}}\right)\, .
\label{surv_diff.2}
\end{equation}

Now, consider our protocol. Let us compute the survival probability  
$Q(t)$ up to time $t$, averaged over the distribution of the starting position at the begining of each cycle.
Then it is easy to see the following.
\begin{itemize}

\item[{\bf a)}] $0<t\le T$: If the measurement time $t$ lies in the first cycle of 
diffusion, then the survival probability upto time $t$ is simply
$Q_1(t)$ given in Eq. (\ref{surv_diff.2}), since no resetting has taken 
place up to $t$ yet. Note that at the end of the period $[0,T]$, the
survival probability is simply $Q_1(T)$.

\item[{\bf b)}] $T<t\le 2T$: In this case, the particle has to first survive the 
period 
$[0,T]$ with free diffusion: this occurs with probability $Q_1(T)$.
Then from time $T$ till $T<t\le 2T$, it also undergoes a free diffusion 
but starting from a new reset position $z$ drawn from ${\cal P}(z)$.
Hence, the survival probability up to $T<t\le 2T$ is given by the product
of these two events
\begin{equation}
Q_2(t)= Q_1(T)\, \int_{-\infty}^{\infty} dz\, {\cal P}(z)\, {\rm
erf}\left(\frac{|L-z|}{\sqrt{4D(t-T)}}\right)\, .
\label{cycle.2}
\end{equation}
Note that at the end of the cycle
\begin{equation}
Q_2(2T)= Q_1^2(T)
\label{cycle_end.2}
\end{equation}
where $Q_1(T)$ is given by Eq. (\ref{surv_diff.2}).

\item[{\bf c)}] $2T<t\le 3T$: By repeating the above argument
\begin{equation}
Q_3(t)= Q_1(2T)\, \int_{-\infty}^{\infty} dz\, {\cal P}(z)\, {\rm
erf}\left(\frac{|L-z|}{\sqrt{4D(t-2T)}}\right)
= Q_1^2(T)\, \int_{-\infty}^{\infty} dz\, {\cal P}(z)\, {\rm
erf}\left(\frac{|L-z|}{\sqrt{4D(t-2T)}}\right)
\label{cycle.3}
\end{equation}
At the end of the 3rd cycle
\begin{equation}
Q_3(3T)= Q_1^3(T)
\label{cycle_end.3}
\end{equation}
where $Q_1(T)$ is given by Eq. (\ref{surv_diff.2}).

\item[{\bf d)}] $(n-1)T<t\le n\,T$: For the $n$-th cycle, we have then
\begin{equation}
Q_n(t)= \left[Q_1(T)\right]^{n-1}\, 
\int_{-\infty}^{\infty} dz\, {\cal P}(z)\, {\rm
erf}\left(\frac{|L-z|}{\sqrt{4D(t- (n-1)\,T)}}\right)    
\label{cycle.n}
\end{equation}

\end{itemize}
\vskip 0.5cm

{Hence, the survival probability $Q(t)$ is just $Q_n(t)$ if $t$ belongs to the $n$-th period, i.e.,
if $(n-1)T<t\le n\,T$, where $n=1,\,2,\,\ldots$. In other words, for fixed $t$, we need to find
the cycle number $n$ associated to $t$ and then use the formula for $Q_n(t)$ in \eqref{cycle.n}.
Mathematically speaking
\begin{equation}
Q(t)= \sum_{n=1}^{\infty} Q_n(t)\, {\Large {\rm I}}_{ (n-1)T< t\le n\, T}
\label{Qt_def}
\end{equation}
where the indicator function ${\Large {\rm I}}_{A}=1$ if the clause $A$ in the subscript
is satisfied and is zero otherwise.}

The mean first-passage time $\langle t_f\rangle$ to location $L$, average over the distribution of $x_0$, is then
given from Eq. (\ref{MFPT.0}) as
\begin{equation}
\langle t_f\rangle = \int_0^{\infty} Q(t) dt
\label{mfpt.1}
\end{equation}
where $Q(t)$ is the survival probability up to time $t$ in \eqref{Qt_def}. Using the
results above for different cycles, we can evaluate this integral
in Eq. (\ref{mfpt.1}) by dividing the time integral over different cycles.
This gives
\begin{eqnarray}
\langle t_f\rangle = \sum_{n=1}^{\infty} \int_{(n-1)T}^{nT} Q_n(t)\, dt
&= & \sum_{n=1}^{\infty} \left[Q_1(T)\right]^{n-1}\,
\int_{(n-1)T}^{nT} dt\, \int_{-\infty}^{\infty} dz\, {\cal P}(z)\, 
{\rm erf}\left(\frac{|L-z|}{\sqrt{4D(t- (n-1)\,T)}}\right) \nonumber \\
&=& \sum_{n=1}^{\infty} \left[Q_1(T)\right]^{n-1}\, 
\int_0^T d\Delta \,
\int_{-\infty}^{\infty} dz\, {\cal P}(z)\,
{\rm erf}\left(\frac{|L-z|}{\sqrt{4D\,\Delta}}\right)\, ,
\label{mfpt.2}
\end{eqnarray}
where in the last line we made a change of variable $\tau= t-(n-1)T$.
The sum on the right hand side (rhs) of Eq. (\ref{mfpt.2}) can be
easily performed as a geometric series. This gives 
\begin{equation}
\langle t_f\rangle = \frac{ \int_0^T d\tau
\int_{-\infty}^{\infty} dz\, {\cal P}(z)\,
{\rm
erf}\left(\frac{|L-z|}{\sqrt{4D\,\tau}}\right)}{1-Q_1(T)}\, .
\label{mfpt.3}
\end{equation} 
The denominator can be further simplified as
\begin{equation}
1-Q_1(T)= 1- \int_{-\infty}^{\infty} dz\, {\cal P}(z)\, {\rm
erf}\left(\frac{|L-z|}{\sqrt{4DT}}\right)
= \int_{-\infty}^{\infty} dz\, {\cal P}(z)\, {\rm
erfc}\left(\frac{|L-z|}{\sqrt{4DT}}\right)
\label{denom.1}
\end{equation}
where ${\rm erfc}(z)= 1-{\rm erf}(z)= (2/\sqrt{\pi})\int_z^{\infty} 
e^{-u^2}\, du$ denotes the complementary error function. Note that
we have used the normalization: $\int_{-\infty}^{\infty} {\cal 
P}(z)\, dz=1$. Substituting the result of Eq. \eqref{denom.1} in
Eq. (\ref{mfpt.3}) we
get our final formula, valid for arbitrary reset/initial 
distribution ${\cal P}(z)$
\begin{equation}
\langle t_f\rangle = \frac{\int_0^T d\tau
\int_{-\infty}^{\infty} dz\, {\cal P}(z)\,
{\rm
erf}\left(\frac{|L-z|}{\sqrt{4D\,\tau}}\right)}{\int_{-\infty}^{\infty} 
dz\, {\cal P}(z)\, {\rm
erfc}\left(\frac{|L-z|}{\sqrt{4DT}}\right)}\, .
\label{mfpt_final}
\end{equation}
The full first-passage probability density $F(t)$, averaged over the distribution of $x_0$, is then given by
\begin{equation}
F(t)= -\frac{dQ}{dt}\, ,
\label{FP.2}
\end{equation}
where $Q(t)$ is given in Eq. (\ref{Qt_def}).


\subsection{ Reset to a Gaussian initial distribution} \label{sec:gaus_periodic}

As an example, let us 
consider the Gaussian distribution for the reset/initial position
\begin{equation}
{\cal P}(x_0) = \frac{1}{\sqrt{2\,\pi\,\sigma^2}}\, e^{- {x_0}^2/{2\sigma^2}}
\label{gauss.1}
\end{equation}
where $\sigma$ denotes the width. In this case, the survival probability $Q_1(T)$ at the end of the first cycle,
is given from Eq. (\ref{surv_diff.2}) upon setting $t=T$
\begin{equation}
Q_1(T)= \int_{-\infty}^{\infty} \frac{dx_0}{\sigma\, \sqrt{2\pi}}\, e^{-{x_0}^2/{2\sigma^2}}\, 
{\rm erf}\left(\frac{L-x_0}{\sqrt{4\, D\, T}}\right)\, .
\label{Q1T_Gauss.1}
\end{equation}
Let us introduce two dimensionless constants $b$ and $c$
\begin{equation}
b= \frac{L}{\sigma}\, ; \quad\quad\quad\quad c=
\frac{L}{\sqrt{4DT}}\, .
\label{bc.1}
\end{equation}
Then in terms of these two constants, Eq. (\ref{Q1T_Gauss.1}), after suitable rescaling, can be simplified to 
\begin{equation}
Q_1(T)= \frac{b}{\sqrt{2\pi}}\, \int_{-\infty}^{\infty} dy\, e^{-b^2\, y^2/2}\, {\rm erf}\left(c\, |1-y|\right)\, .
\label{Q1T_Gauss.2}
\end{equation}
Note that in the limit of $b\to \infty$, i.e., when $L\ll \sigma$ (corresponding to resetting to
the origin) one gets 
\begin{equation}
Q_1(T)\Big |_{b\to \infty}= {\rm erf}(c) \, ,
\label{Q_1.blarge.1}
\end{equation}
where we used $ (b/\sqrt{2\pi})\, e^{-b^2\, y^2/2} \to \delta(y)$ when $b\to \infty$. 
 
\subsubsection{ Mean first-passage time}
Substituting the Gaussian 
${\cal P}(z)$ in Eq. (\ref{mfpt_final}), and rescaling $\tau= v\, T$ we 
can write everything in dimensionless form
\begin{equation}
\tau=\frac{4D\langle t_f\rangle}{L^2} = \frac{\int_0^1 dv
\int_{-\infty}^{\infty} du\, e^{-u^2/2}\,
{\rm
erf}\left(   
\frac{c}{\sqrt{v}}\,|1-u/b|
\right)}{c^2\, \int_{-\infty}^{\infty} du\, 
e^{-u^2/2}\, 
{\rm erfc}\left(c\,|1-u/b|\right)}\, \equiv w(b,c)
\label{mfpt_gauss.1}
\end{equation}
where the dimensionless constants $b$ and $c$ are given in Eq. (\ref{bc.1}).
It is hard to obtain a more explicit expression for the function $w(b,c)$ in
Eq. (\ref{mfpt_gauss.1}). But it can be easily evaluated numerically.

To make further analytical progress, we first consider the limit 
$\sigma\to 0$, i.e., $b=L/\sigma\to \infty$. 
In this limit, the reset distribution
${\cal P}(z)\to \delta(z)$. Hence, Eq. (\ref{mfpt_final}) 
or equivalently Eq. (\ref{mfpt_gauss.1}) simplifies 
considerably and the integrals can be evaluated
explicitly. We obtain an exact 
expression
\begin{equation}
\tau= \frac{4D\langle t_f\rangle}{L^2} = \frac{{\rm 
erf}(c)+2c\left(e^{-c^2}/\sqrt{\pi}-c\, {\rm erfc}(c)\right)}{c^2\, 
{\rm erfc}(c)}\equiv w(b\to \infty, c)\equiv w(c)
\label{delta_reset.1}
\end{equation}
where we recall $c= L/\sqrt{4DT}$. In Fig. 2 of the main text, we plot the 
function 
$w(c)$
vs. c, which  has a  minimum at 
\begin{equation}
c_{\rm opt}= 0.738412...
\label{copt.1}
\end{equation}
At this optimal value, $\tau_{\rm opt}= w(c_{\rm opt})= 5.34354\ldots$.
Hence, the optimal mean first-passage time to find the target located at 
$L$ is given by
\begin{equation}
\langle t_f\rangle_{\rm opt}= (5.34354\ldots)\, \frac{L^2}{4D}\, .
\label{tfopt.1}
\end{equation}
Note that this result is true in the limit $b=L/\sigma\to \infty$, i.e., 
when the target is very far away from the starting/resetting position.
In this limit, our result coincides with Ref. \cite{PKE16} where 
the authors studied
periodic resetting to the fixed initial position $x_0=0$ by a different method.
This is expected since the limit $b=L/\sigma \to \infty$ limit
is equivalent to $\sigma\to 0$ (with fixed $L$) and one would expect to
recover the fixed initial position results.

But the most interesting and unexpected result occurs for finite $b$.
In this case we can evaluate the rhs of Eq. 
(\ref{mfpt_gauss.1}) numerically. This has been done to plot the continuous lines 
in Fig. 3 of the main text, which shows the existence of a metastable minimum 
for $b>b_c\simeq 2.3$

\subsubsection{Full first-passage probability density $F(t)$}

The first-passage probability density can be computed from Eq. (\ref{FP.2}) and is plotted in Fig. 2 of the main text.
For a finite $b=L/\sigma$ we see those spectacular spikes at the begining of each cycle. 
In this subsection, we analyse the origin of these spikes. For this let us analyse the survival probability
$Q_n(t)$ in Eq. (\ref{cycle.n}) close to the epoch $n\, T$, i.e., when the $n$-th period ends.
We set $t= n\, T+\Delta$ where $\Delta$ is small. For $\Delta>0$, we are in the $(n+1)$-th cycle, while for $\Delta<0$
we are in the $n$-th cycle. We consider the $\Delta>0$ and $\Delta<0$ cases separrately.

\vskip 0.3cm

{\noindent {\bf {The case $\Delta>0$:}}} Since $t= n\, T+\Delta$ with $\Delta>0$ small,
$t$ now belongs to the $(n+1)$-th cycle, so we replace $n$ by $n+1$ in Eq. (\ref{cycle.n}) and get
\begin{equation}
Q_{n+1}(t= n\, T+ \Delta)= \left[Q_1(T)\right]^n\, \int_{-\infty}^{\infty} 
\frac{dz}{\sigma\, \sqrt{2\pi}}\, e^{-z^2/{2\sigma^2}}\,
{\rm erf}\left(\frac{|L-z|}{\sqrt{4\, D\, \Delta}}\right)\, ,
\label{right_cycle_n.1}
\end{equation}
where $Q_1(T)$ is given in Eq. (\ref{Q1T_Gauss.2}). It is convenient first to use the relation
${\rm erf}(z)= 1-{\rm erfc}(z)$ and rewrite this as
\begin{equation}
Q_{n+1}(t= n\, T+ \Delta)= \left[Q_1(T)\right]^n\, \left[1- \int_{-\infty}^{\infty}
\frac{dz}{\sigma\, \sqrt{2\,\pi}}\, e^{-z^2/{2\sigma^2}}\,
{\rm erfc}\left(\frac{|L-z|}{\sqrt{4\, D\, \Delta}}\right)\,\right]\,  
\label{right_cycle_n.2}
\end{equation}
To derive the small $\Delta$ asymptotics, we make a change of variable $(L-z)/\sqrt{4\,D\,\Delta}=y$ on the rhs
of Eq. (\ref{right_cycle_n.2}). This leads to, using $b=L/\sigma$,
\begin{equation}
Q_{n+1}(t= n\, T+ \Delta)= \left[Q_1(T)\right]^n\, \left[1-\frac{\sqrt{4\,D\, \Delta}}{L}\, \frac{b}{\sqrt{2\,\pi}}\,
\int_{-\infty}^{\infty} dy\, e^{-\frac{b^2}{2}\, (1- \sqrt{4\,D\,\Delta}\, y/L)^2}\, {\rm erfc}(|y|)\, \right]\, .
\label{right_cycle_n.3}
\end{equation}
We can now make the limit $\Delta \to 0$. To leading order in $\Delta$, we can set $\Delta=0$ inside the exponential
in the integrand on the rhs and using $\int_{-\infty}^{\infty} dy\, {\rm erfc}(|y|)=2/\sqrt{\pi}$ we get
\begin{equation}
Q_{n+1}(t= n\, T+ \Delta)\simeq \left[Q_1(T)\right]^n\, \left[1- A_{+}(b)\, \sqrt{\Delta}\, \right]\, , \quad\quad
A_+(b)= \sqrt{\frac{8\,D}{\pi^2\, L^2}}\, b\, e^{-b^2/2}\, . 
\label{right_cycle_n.4}
\end{equation}
Taking derivative with respect to time, i.e., with respect $\Delta$, we find that as $\Delta\to 0^+$,
the first-passage probability denity $F(t)=-dQ/dt$ diverges for any finite $b$ as 
\begin{equation}
F(t= n\, T+ \Delta) \simeq \frac{A_n(b)}{ \sqrt{\Delta}}
\label{fp_n+.1}
\end{equation}
where the amplitude $A_n(b)$ of the inverse square root divergence is given by the exact formula
\begin{equation}
A_n(b)= \sqrt{\frac{2D}{\pi^2\, L^2}}\, b\, e^{-b^2/2}\, \left[Q_1(T)\right]^{n}\, 
\label{Ab.1}
\end{equation}
where $Q_1(T)$ (which also depends on $b$) is given in Eq. (\ref{Q1T_Gauss.2}). This inverse square root 
divergence at the
begining of any cycle, for any finite b,  describes the spikes in Fig. 2 of the main text. 
Note that the amplitude $A_n(b)$
of the $n$-th spike in Eq. (\ref{Ab.1}) decreases exponentially with $n$, as seen also in Fig. 2 of the main text.

Interestingly, we see from Eq. (\ref{Ab.1}) that the amplitudes of the spikes
decrease extremely fast as $b$ increases and the spikes disappear in the $b\to \infty$ limit.
Since $b=L/\sigma$, we see that the limit $b\to \infty$ corresponds to $\sigma\to 0$ limit (for fixed $L$).
This is indeed the case where one resets always to the origin. In fact, if we first take the limit $b\to \infty$
keeping $\Delta$ fixed and then take the limit $\Delta\to 0$, we get a very different reesult. Taking $b\to \infty$
limit first in Eq. (\ref{right_cycle_n.3}) we get, 
\begin{equation}
Q_{n+1}(t= n\, T+ \Delta)\Big|_{b\to \infty}= \left[Q_1(T)\right]^n 
\left[1- {\rm erfc}\left(\frac{L}{\sqrt{4\,D\,\Delta}}\right)\right]\, ,
\label{Q+blarge_n.1}
\end{equation}
where $Q_1(T)$ now is given in Eq. (\ref{Q_1.blarge.1}). Taking a derivative with respect $t$, i.e.,
with respect to $\Delta$, 
we get the first-passage probability density
\begin{equation}
F(t= n\, T+ \Delta)\Big|_{b\to \infty}= \left[Q_1(T)\right]^n\, \frac{L}{\sqrt{4\, \pi\, D\, \Delta^3}}\, 
e^{-L^2/(4\,D\,\Delta)}\, .
\label{fp+blarge_n.1}
\end{equation}
Thus, in the $b\to \infty$ limit, the first pasage probability actually vanishes extremely rapidly as $\Delta\to 0$
due to the essential singular term $\sim e^{-L^2/(4\, D \, \Delta)}$.

To summarize, the behavior of the first-passage probability density at the begining of the $n$-th 
period is strikingly different for $\sigma=0$ ($b\to \infty)$
and $\sigma>0$ ($b$ finite): in the former case it vanishes extremely rapidly as $t\to n\, T$ from above, 
while in the latter case it diverges as an inverse sqaure root leading to a spike at the begining of each period.
Thus the occurrence of spikes for $\sigma>0$ is a very clear signature of the finiteness of $\sigma$.
Physically, a spike occurs because if $\sigma$ is finite, immedaitely after each resetting the searcher
may be very close to the target and has a finite probability of finding the target immediately without further
diffusion.

\vskip 0.3cm

{\noindent {\bf {The case $\Delta<0$:}}} When $t= n\, T+ \Delta$ with $\Delta<0$, the effects are not so prominent
as in the $\Delta>0$ case. For $\Delta<0$, since $t$ belongs to the $n$-th branch, we consider the formula
in Eq. (\ref{cycle.n}) with $n$ and get
\begin{equation}
Q_{n}(t= n\, T+ \Delta)= \left[Q_1(T)\right]^{n-1}\, \int_{-\infty}^{\infty}
\frac{dz}{\sigma\, \sqrt{2\pi}}\, e^{-z^2/{2\sigma^2}}\,
{\rm erf}\left(\frac{|L-z|}{\sqrt{4\, D\, (T+\Delta)}}\right)\, .
\label{left_cycle_n.1}
\end{equation}
Now taking the $\Delta\to 0$ limit in Eq. (\ref{left_cycle_n.1}) is straightforward since there is no singular
behavior. Just Taylor expanding the error function for small $\Delta$ up to $O(\Delta)$, and making the
change of variable $(L-z)/\sqrt{4\, D\, T}=y$,  we obtain
\begin{equation}
Q_{n}(t= n\, T+ \Delta)= \left[Q_1(T)\right]^{n} - \Delta\, B_n(b) + O(\Delta^2)\, ,
\label{left_cycle_n.2}
\end{equation}
where the constant
\begin{equation}
B_n(b)= \frac{b \,\left[Q_1(T)\right]^{n-1}}{c\, T\, \pi\, \sqrt{2}}\,
\int_{-\infty}^{\infty} dy\, |y| \, e^{-y^2} \, e^{-\frac{b^2}{2}\,(1-c\, y)^2}\,  ,
\label{B_nb.1}
\end{equation}
with $c=L/\sqrt{4\, D\, T}$. 
Taking derivative with respect to $\Delta$ in Eq. (\ref{left_cycle_n.2}) 
gives the leading behavior of the first-passage probability density: it approaches
a constant as $\Delta\to 0$
\begin{equation}
F(t= n\, T+ \Delta) \simeq B_{n}(b) \, .
\label{left_fp.1}
\end{equation}
One can easily check that even in the $b\to \infty$ limit, this constant remains nonzero and is given by
\begin{equation}
F(t= n\, T+ \Delta)\Big|_{b\to \infty} \simeq \frac{\left[Q_1(T)\right]^{n-1}}{\sqrt{\pi}\, 
T\, c^3}\, e^{-1/(2\, c^2)}\, ,
\label{left_fp_blarge.1}
\end{equation}
where $Q_1(T)$ is now given in Eq. (\ref{Q_1.blarge.1}).
Thus, for any $\sigma\ge 0$, as one approaches
the epoch $t=n\, T$ from the left, i.e., at the end of the $(n-1)$-th cycle, 
the first-passage probability density approaches a constant $B_n(b)$ given
in Eq. (\ref{B_nb.1}). {This constant remains finite even when $b\to \infty$.
On this side, there is no spectacular
effect distinguishing $\sigma=0$ ($b\to \infty$) and $\sigma>0$ (finite $b$), as is the 
case on the right side of the epoch $n\, T$.
On the right, for finite $b$ we have an inverse square root divergence as in \eqref{fp_n+.1}, while
for $b\to \infty$, it vanishes rapidly as $\Delta\to 0$. Thus in the large $b$ limit, we have
a discontuity or drop from left to right (not a spike) as we see in the inset of Fig. 2
in the main text.}    

\section{Protocol-2: Random resetting to  a random initial position}

We consider an overdamped diffusing particle that starts at an initial
position $x_0$, which is drawn from a distribution ${\cal P}(x_0)$.
The particle diffuses for a random interval $T$ drawn from an 
exponential distribution $P(T)= r\, e^{-r\,T}$  and then its position is
instantaneously reset to a new position $z$, also drawn from the same
initial distribution ${\cal P}(z)$. Then the particle diffuses again 
for another random interval
$T$ (again drawn from the exaponential distribution $P(T)$), 
followed by a reset to a new position $z'$ drawn from ${\cal P}(z)$
and the process continues. We assume that after each resetting, the reset 
position $z$ is drawn {\rm independently} from cycle to cyle from the
same distribution ${\cal P}(z)$. Similarly, the interval $T$ is
also drawn independently from cycle to cyle from the same distribution
$P(T)= r\, e^{-r\, T}$.
We also have a fixed target at a
location $L$. For fixed $L$, $r$ and ${\cal P}(z)$, We want to 
first compute the mean first-passage time
$\langle t_f\rangle$ to find the target and then optimize (minimize) this
quantity with respect to $r$ (for fixed $L$ and ${\cal P}(z)$). Recall
that in protocol-1, we just had a fixed $T$, but in protocol-2
$T$ is exponentially distributed.

\vskip 0.4cm

In this case, the mean first-passage time $\langle t_f\rangle$ has a
simple closed formula. This formula can be derived folllowing
steps similar to those in Ref. \cite{EM2011_2}. Skipping further
details, we find
\begin{equation}
\langle t_f\rangle= \frac{1}{r}\,\left[\frac{1}{\int_{-\infty}^{\infty}
	dz\, {\cal P}(z)\, \exp\left(- \sqrt{\frac{r}{D}}\, |L-z|\right)}-1\right]\, .
\label{mfpt.1p2}
\end{equation}

\vskip 0.4cm

Consider the special case of Gaussian distribution of Eq. (\ref{gauss.1})
Substituting this Gaussian 
${\cal P}(z)$ in Eq. (\ref{mfpt.1p2}), and performing the integral explicitly, we get
the dimensionless mean first-passage time
\begin{equation}
\tau=\frac{4D\langle t_f\rangle}{L^2} = 
\frac{4}{c^2}\, \left[\frac{2\,e^{-c^2/{2b^2}}}{e^c \, 
	{\rm erfc}\left(\frac{1}{\sqrt{2}}\left(\frac{c}{b}+b\right)\right)
	+ e^{-c} \, {\rm erfc}\left(\frac{1}{\sqrt{2}}\left(\frac{c}{b}-b\right)\right)}-1\right]\, \equiv w_2(b,c)
\label{mfpt.2p2}
\end{equation}
where the dimensionless constants $b$ and $c$
are given by
\begin{equation}
b= \frac{L}{\sigma}\, ; \quad\quad\quad\quad c= 
\sqrt{\frac{r}{D}}\, L\, .
\label{bc.1p2}
\end{equation}
Recall that in periodic protocol (with a fixed $T$ discussed in section \ref{sec:gaus_periodic}) , the parameter $b=L/\sigma$ is the same, but the
parameter $c=L/\sqrt{4DT}$ is  different.
In protocol-2, the reset rate $r$ effectively plays the role of $1/T$ in protocol-1.

\vskip 0.4cm

The formula for $w_2(b,c)$ in Eq. (\ref{mfpt.2p2}) for protocol-2 is simpler and more explicit, 
compared to the equivalent formula Eq.\ref{mfpt_gauss.1} for protocol-1.   

\vskip 0.4cm

 In Eq. (\ref{mfpt.2p2}), consider first the limit $b\to \infty$, i.e.,
resetting to a fixed initial condition. In this case, Eq. (\ref{mfpt.2p2}) reduces precisely to the
formula derived by Evans and Majumdar~\cite{EM2011}
\begin{equation}
\tau= w_2(b\to \infty, c)= \frac{4}{c^2}\, \left[e^c-1\right]\, .
\label{b_infty.p2}
\end{equation}
In this $b\to \infty$ limit, $\tau$ as a function of $c$, has a unique minimum at $c=c^*=1.59362\ldots$,
where $\tau(c=c^*)=6.17655\ldots$.

\vskip 0.4cm

For finite fixed $b$, it is very easy to plot the function $w_2(b,c)$ vs. $c$ given in
Eq. (\ref{mfpt.2p2}) and one finds
that as in protocol-1, $w_2(b,c)$ first decreases with increasing $c$, achieves a minimum
at certain $c_1(b)$, then starts increasing again--becomes a maximum at $c_2(b)$ and finally decreases
for large $c$ as $\sim 1/c$ (this asymptotics for large $c$ can be computed exactly from
Eq. (\ref{mfpt.2p2})). Note that the actual values of $c_1(b)$band $c_2(b)$ are of course different in
the two protocols. Interestingly, this metastable optimum time at $c=c_1(b)$ 
disappears when $b$ becomes smaller
than a critical value $b_c= 2.53884$. For $b<b_c$, the function $w_2(b,c)$ decreases
monotonically with increasing $c$. Hence, for $b<b_c$, the best solution is $c=\infty$ (i.e.,
repeated resetting). The curves plotted in Fig. 4 of the main text correspond to this theoretical prediction
 and they have been checked experimentally and numerically. This numerical test allows us to clearly 
state that the small disagreement at small $b$ and large $c$ of the experimental data  
with the theoretical predictions is due to the maximum sampling rate of our analog to digital 
converter which was not fast enough to detect very short MFPT.